\documentclass[12pt]{article}
\usepackage{eucal,  amssymb,amsmath,graphicx, amsfonts, latexsym, epsfig, multirow, hhline}
\usepackage[T1]{fontenc}

\begin{document} \parskip=5pt plus1pt minus1pt \parindent=0pt
\title{Inferring global network properties from egocentric data with applications to epidemics}
\author{Tom Britton\thanks{ {\it Corresponding author}: Department of
Mathematics, Stockholm University, SE-106 91 Stockholm, Sweden. {\it
E-mail}: tom.britton@math.su.se, {\it Phone:} +46 8 164534}  and Pieter Trapman\thanks{Department of
Mathematics, Stockholm University, SE-106 91 Stockholm, Sweden.}}
\date{\today}
\maketitle

\begin{abstract}
Social networks are rarely observed in full detail. In many situations properties are known for only a sample of the individuals in the network and it is desirable to induce global properties of the full social network from this ''egocentric'' network data. In the current paper we study a few different types of egocentric data, and show what global network properties are consistent with those egocentric data. Two global network properties are considered: the size of the largest connected component in the network (the giant), and secondly, the possible size of an epidemic outbreak taking place on the network, in which transmission occurs only between network neighbours, and with probability $p$. The main conclusion is that in most cases, egocentric data allow  for a large range of possible sizes of the giant and the outbreak. However, there is an upper bound for the latter. For the case that the network is selected uniformly among networks with prescribed egocentric data (satisfying some conditions), the asymptotic size of the giant and the outbreak is characterised.

\vskip1cm
\end{abstract}


\section{Introduction}
Social network data may be of different levels of detail, e.g.\ complete (sociocentric), snowball sampled, egocentric with alter connections or ego-only egocentric \cite{HR}. Here egocentric data means that information of the immediate surrounding of a sample of actors are collected. More precisely, following Hanneman and Riddle, we distinguish between  ego-only egocentric data where the connections of each sampled actor (ego) is all that is collected, and egocentric with alter connections, where it is also observed which of these connections are themselves connected. To observe the complete network in a large community is of course expensive and time consuming, which is the reason why data often consists of snowball samples or egocentric data (e.g.\ \cite{HR}). Clearly, the higher level of detail in the collected data, the more can be inferred with higher precision \cite{MM}. However, as has been shown by Marsden \cite{M02} who studies betweenness, it is in some situations possible to infer also global network properties from egocentric data in a fairly robust way. Many social networks share the property known as transitivity (closely related to clustering) that if $A$ is connected to $B$ and $B$ is connected to $C$, then it is more likely that $A$ is also connected to $C$ (e.g.\ \cite{KHRH}). To which extent this property is manifested in a given social network is obviously better known from egocentric data with alter connections as compared with ego-only egocentric data. As a consequence, global properties affected by transitivity/clustering should be easier to infer from the former type of data.

In the current paper we investigate what can be deduced about global network properties when observing different sorts of egocentric data. More precisely, we focus on the (relative) size $\tau$ of the largest connected component of the network (the giant) when all that is known is the mean degree of actors, when ego-only egocentric data are collected, and where egocentric data with alter connections are observed. Additional to this we study possibly scenarios for an epidemic spreading ''on'' the social network. The main conclusion is that very little can be said about $\tau$ if only egocentric data of any type without additional information are observed. The same is however not true for epidemics occurring on the (more specified) social network: the more detailed information about the egocentric network the narrower is the range of possible outbreak sizes. We also study the size of the connected component and the epidemic outbreak size of a \emph{random} or \emph{typical} network with the prescribed egocentric properties.

\section{Network properties and epidemic model}

Consider a community, or social network, consisting of $n$ individuals/actors, where $n$ is assumed to be large. Each pair of actors $i$ and $j$ are either connected by an (undirected) edge or not, where the edge reflects some type of social relationship (liking, shared membership of group or household, sexual relationship, ...). Let $d_{i,j}=d_{j,i}=1$ if $i$ and $j$ are connected and $d_{i,j}=d_{j,i}=0$ otherwise. Knowing $d_{i,j}$ for all $i$ and $j$ then corresponds to knowing the complete network. Knowing ego-only egocentric data means that we only know $d_i=\sum_jd_{i,j}$, the number of connections, or the degree, of all or a sample of the actors. This means that we know the degree \emph{distribution} in the community. Below we will also study the situation where we have even less information, i.e.\ where all that is observed is the \emph{mean} degree $\mu_D=\sum_id_i/n$. Thus, in the first situation we know what fraction $p_0$ that has degree 0, what fraction $p_1$ that has degree 1 and so on (i.e.\ we know the degree distribution $\{p_k\}$), whereas in the latter case we only know that the mean degree equals a certain number $\mu_D$. Knowing the degree distribution will also give the mean degree by the relation $\mu_D=\sum_kkp_k$.

\begin{figure}[ht]
\begin{center}
\includegraphics[width=0.3\textwidth]{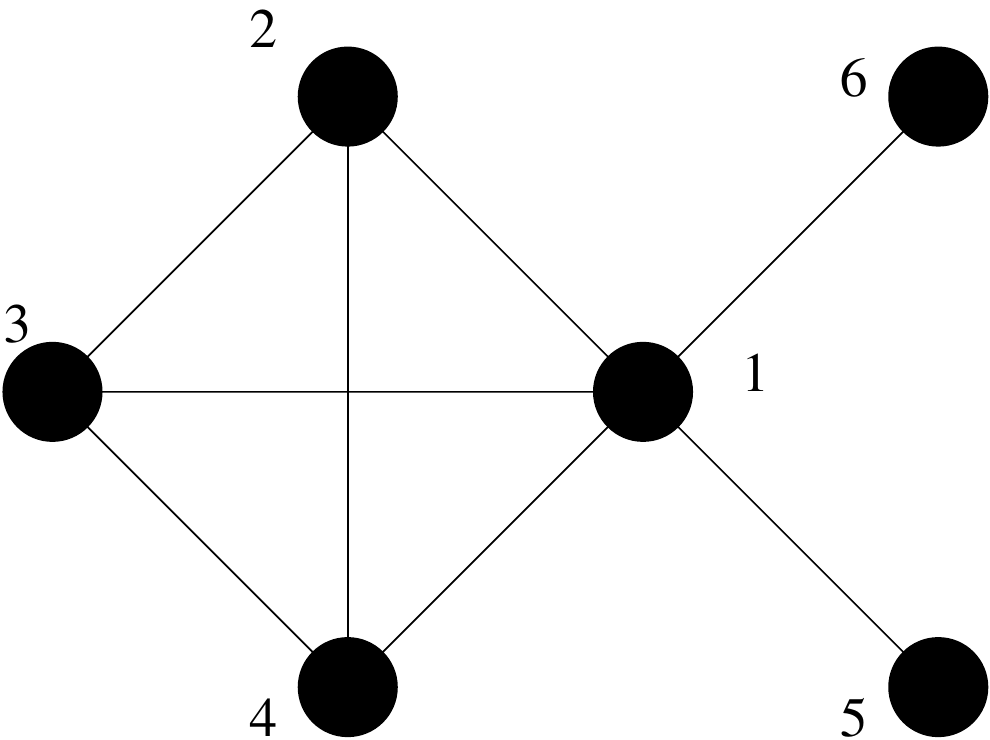}
\end{center}
\caption{A mini graph in which actor 1 has single degree 2 and triangle degree 3}\label{minigraph}
\end{figure}

If we have egocentric data with alter connections we also know which connections of an actor are connected themselves. That is, if $d_{i,j}=1$ and $d_{i,k}=1$ we observe whether $d_{j,k}=1$ or not. Later we will simplify this type of data to the situation that we specify two degrees of each actor: the single degree and the triangle degree, where the single degree is the number of actors ''ego'' is connected to that are not connected to any other acquaintance of ego, and the triangle degree denotes how many triangles ego is part of \cite{M09}. So for example, in Figure \ref{minigraph} ego (actor 1) is connected to 5 actors, 3 which together with ego all know each other and 2 separate actors that, each of them, don't know anyone else of ego's neighbours. So, ego has single degree 2 and triangle degree 3 (there are three ways to chose 2 out the 3 common friends) 
Admittedly, by reducing the egocentric data with alter connections to only keeping track of actors single and triangle degree we loose some information. The reason for doing this is that it makes the mathematical analysis more tractable and it is our hope that it will have only minor effect on the results. To conclude, we represent the egocentric data with alter connections by the single and triangle degree distribution $\{p_{k_1,k_{\Delta}}\}$, where $p_{k_1,k_{\Delta}}$ denotes the fraction of actors that have single degree $k_1$ and triangle degree $k_{\Delta}$.

The first global property we investigate is the relative size of the largest connected component $\tau$. Individuals in the network are said to be directly connected if there is an edge between the actors, and in general nodes are said to be connected if there is a path of directly connected actors between them. The network can hence be decomposed into separate connected components, and the largest (in terms of number of actors) of these components is called the giant component. The relative size $\tau$ is the size (the number of actors) of the giant divided by the population size.

As mentioned earlier we will also see what the effect of an infectious disease spreading in the community, i.e.\ ''on'' the social network, is, in the sense that an infected actor may infect any of its (not yet infected) connected actors but no one else. We assume that the epidemic is initiated by one uniformly at random selected index case, and that anyone that gets infected infects each of its susceptible neighbours independently with probability $p$ (those who get infected may spread the disease to their not yet infected neighbours, and so on). This model is known as the Reed-Frost epidemic model on a network. It is well-known that, for such epidemics taking place in a large community, two qualitatively different things may happen. Either only few actors will get infected (a minor outbreak) or else a positive (hardly random) fraction will get infected; we say a major outbreak has occurred (e.g.\ \cite{N03,B10}). Another known fact for these class of models is that the \emph{probability} $\pi$ of observing a major outbreak equals the relative \emph{size} $\tau^{(epi)}$ of the major outbreak, i.e.\ $\pi=\tau^{(epi)}$. In the current paper we are mainly interested in the relative size of the epidemic, but for the reason just mentioned we may equally well compute the probability $\pi$ of having a major epidemic outbreak. The case $p=1$ implies that the disease spreads to the whole connected component of the index case. The probability $\pi$ of a major outbreak is then equal to the probability that the index case belongs to the giant component, and because the index case is chosen uniformly at random this is the same as the relative size $\tau$ of the giant. We hence have $\tau^{(epi)}=\tau$ when $p=1$.

In the next sections we investigate what the range of possible relative sizes of the giant connected component is, assuming that certain local features of the social network are given. We also study what effect an epidemic taking place on the social network may have. We start by only assuming that the mean degree $\mu_D$ is known, and then gradually assume more informative egocentric data.

\section{Observing only mean degree}\label{sec-mean-deg}

Suppose first that all we know about the network is that mean degree equals $\mu_D$ ($0<\mu_D<\infty$). First we study properties of the largest connected component, and then the size of an epidemic outbreak occurring on the social network.

\subsection{The giant connected component}

What might the relative size of the giant connected component $\tau$ be? Since very little is fixed (only the mean degree $\mu_D$) we may choose rather freely in order to maximise/minimise $\tau$. If we want to minimise $\tau$ we simply make small fully connected and isolated components of size $\lfloor \mu_D\rfloor$ and $\lceil \mu_D\rceil$ where $\lfloor \mu_D\rfloor$  is the integer part of $\mu_D$ and $\lceil \mu_D\rceil$ is the smallest integer, which is larger than or equal to $\mu_D$. If $\mu_D$ is an integer, say $\mu_D=5$, we simply group actors into fully connected groups of size 6 (actors then have degree 5). As a consequence, all connected components have size 6 and the relative size of the largest connected component is $6/n\approx 0$ implying that \[
\tau_{\min}=0.
\]

If we instead want to maximise $\tau$, this is achieved (among other ways) by connecting all actors with degree 2 or more into one giant component (first make a ''line'' out of all actors and then add edges arbitrarily). We hence want to maximise the fraction having degree 2 or more. If $\mu_D\ge 2$ all actors can have degree larger than or equal to 2, which implies that all actors may be connected into one giant component, i.e.\ $\tau_{\max}=1$. If $\mu_D<2$ we have to ''sacrifice'' a fraction of the actors such that the remaining actors all have degree 2. (We can connect two degree 1 actors to the end points of the line, but since we assume that the population is large, those two actors have only marginal effect on the size of the giant.) More specifically, we let a fraction $1-\mu_D/2$ have degree 0 and the remaining fraction $\mu_D/2$ have degree two and putting the latter in one long line. The size of the largest connected component then equals $\tau_{\max}=\mu_D/2$. To conclude, if $\mu_D\ge 2$ then $\tau_{\max}=1$ and otherwise $\tau_{\max}=\mu_D/2$. 

This feature, that $\tau_{\min}=0$ and $\tau_{\max}=1$ or close to 1 (if the mean degree is large enough) will be repeated also when we observe the ego-only egocentric network or with alter connections.

We now pick a network at random among all networks having mean degree $\mu_D$. Having the mean degree fixed and equal to $\mu_D$ is identical to having the total \emph{number} of edges constant and equal to $n\mu_D/2$ (the denominator 2 comes from the fact that each edge contributes to the degree of two different nodes). Choosing a network with $n$ nodes and $m=n\mu_D/2$ edges uniformly at random is a well-known model of Erd\H{o}s-R{\'e}nyi denoted $G(n, m)$ and it is well-known that this network is (for our purposes) asymptotically equivalent to the more familiar $G(n,p=\mu_D/n)$ network of  Erd\H{o}s and R{\'e}nyi in which edges appear between different pairs of nodes independently, each with probability $\mu_D/n$ \cite{Bol01,D06}.

\begin{figure}[ht]
\begin{center}
\includegraphics[width=0.4\textwidth]{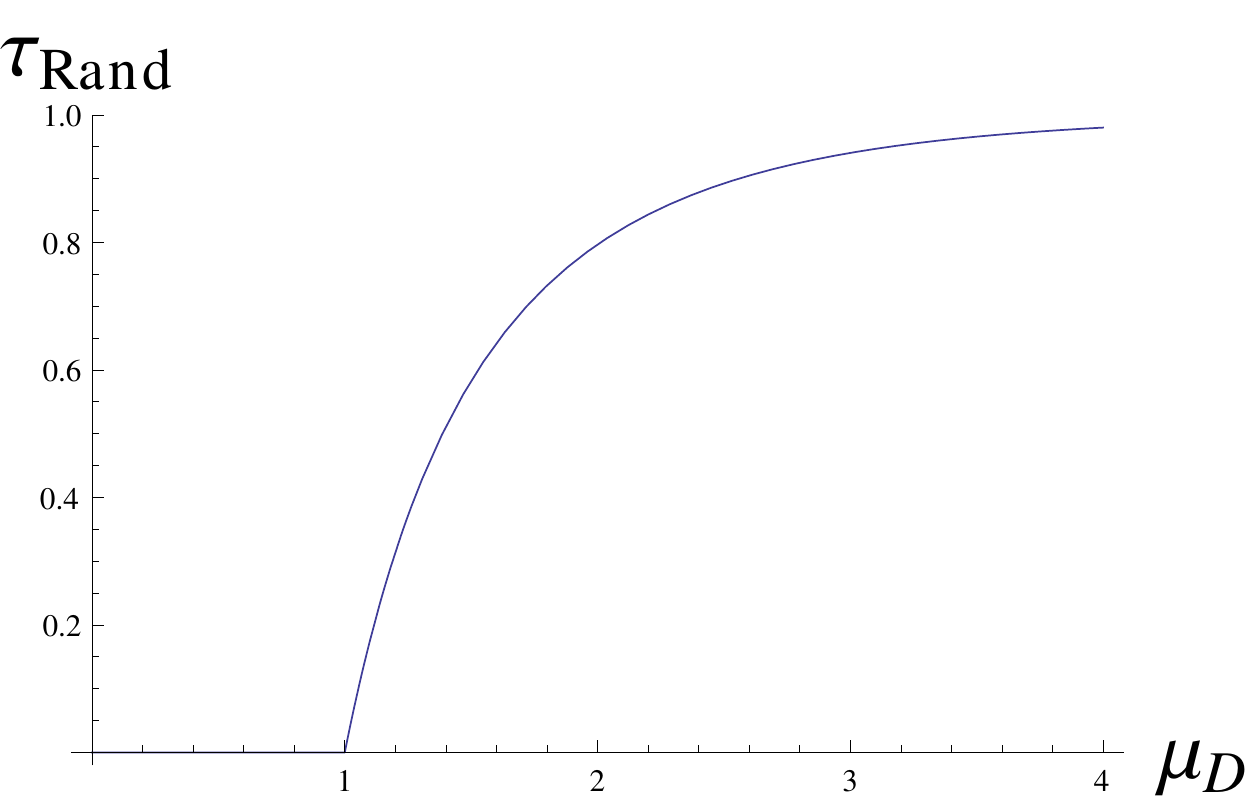}
\end{center}
\caption{The relative size $\tau_{Rand}$ of the largest connected component in an Erd\H{o}s-R{\'e}nyi network.}\label{fig-tau-rand}
\end{figure}

The relative size $\tau_{Rand}$ of the largest connected component in this network is given by the largest solution $\tau_{Rand}=t$ to the equation
\begin{equation}
1-t=e^{-\mu_Dt}.\label{tau-rand-mean}
\end{equation}
See Figure \ref{fig-tau-rand} for an illustration of how $\tau_{Rand}=t$ depends on $\mu_D$.
It is also known that $\tau_{Rand}$ is strictly positive if and only if $\mu_D>1$ \cite{ER,Bol01,D06}.

\subsection{Epidemic outbreak size}

Suppose now that we are interested in the potential spread of an infectious disease taking place on the social network having mean degree $\mu_D$. More precisely, we assume that the transmission model is as defined above, with transmission probability $p$ ($0<p<1$), and that the mean degree $\mu_D$ is all that is given about the social network.

How big can a major outbreak be? Having transmission probability $p$ means that we erase each existing edge with probability $1-p$ and keep it with probability $p$. This will have the effect of possibly breaking up the original largest connected component, but never making it bigger. As a consequence, we still have
\[
\tau^{(epi)}_{\min } =0,
\]
since it was shown in the previous section that $\tau_{\min}=0$.

\begin{figure}[ht]
\begin{center}
\includegraphics[width=0.5\textwidth]{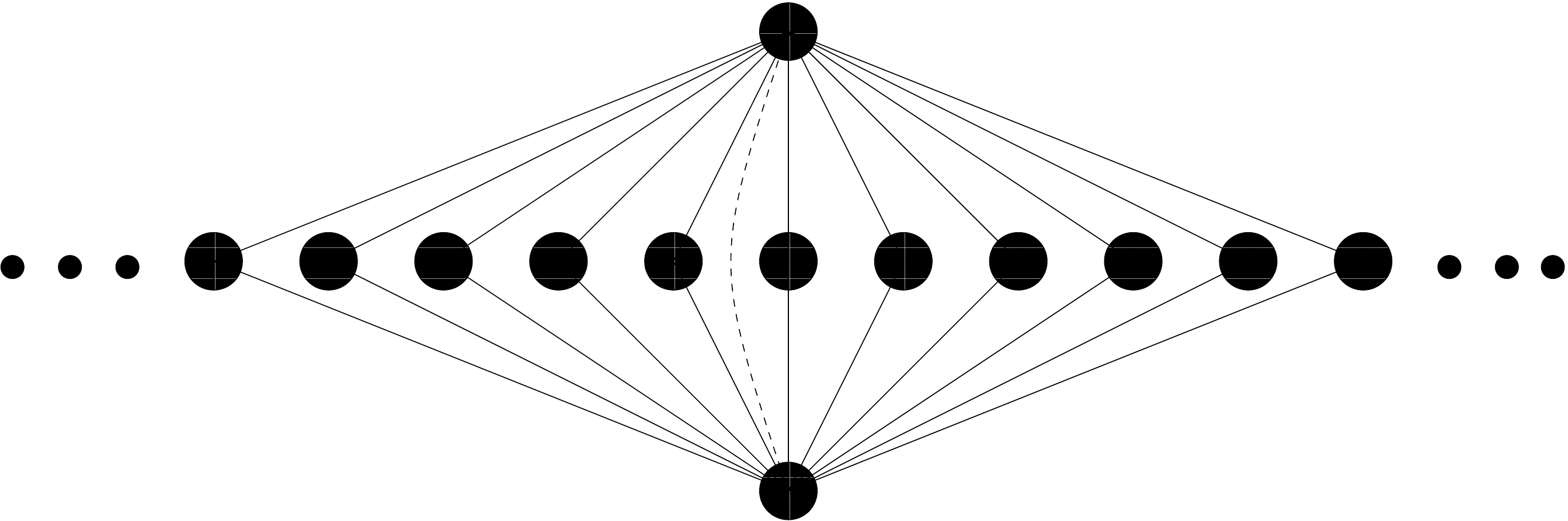}
\end{center}
\caption{Illustration of a large network having mean degree $\mu_D=4$ that maximises the probability and size of a major outbreak. The relative outbreak size in case of a major outbreak equals $\tau^{(epi)}_{\max} = 1-(1-p)^2$ where $p$ is the transmission probability.}
\label{fig-star}
\end{figure}

Because the epidemic has the effect of removing some edges (often denoted thinning) it is very unlikely that everyone gets infected. How do we maximise the size of a major outbreak for given $\mu_D$ and transmission probability $p$? Let us first consider the case where $\mu_D$ is a multiple of 2. One choice of network that maximises the outbreak size/probability $\tau^{(epi)}$ is then to let $\mu_D/2$ (an integer) number of actors each be connected to \emph{every} other actor (we call them central nodes), and the remaining $n-\mu_D/2$ actors each only being connected to these central actors (see Figure \ref{fig-star} for an illustration of this ''starlike'' construction for the case $\mu_D=4$). The mean degree of this network equals
\[\frac{\mu_D/2}{n}(n-1) + \frac{n-\mu_D/2}{n}\mu_D/2\approx \mu_D,\]
the approximation relying on $n$ to be large.

To compute the probability $\pi$ of a large epidemic outbreak for this network is straightforward, and as before we have $\pi=\tau^{(epi)}$, the relative size of a major outbreak. The index case is selected randomly; most likely it is hence one of the nodes having degree $\mu_D/2$. However, if this actor infects at least one of its neighbours, then a major outbreak will certainly occur since all of its neighbours are central actors, each with degree $n-1$. The probability that the actor infects at least one neighbour is $1-(1-p)^{\mu_D/2}$ which hence equals $\pi=\tau^{(epi)}$. This reasoning is easily extended to the case that $\mu_D/2$ is not an integer.
To this end let $\mu_D/2= \lfloor \mu_D/2\rfloor +\alpha$ where $\lfloor \mu_D/2\rfloor$ is the integer-part of $\mu_D/2$ and $\alpha$ the remainder. Then there should be $\lfloor \mu_D/2\rfloor$ central nodes, each  connected to all other nodes, and one node connected to $\alpha n$ other nodes (if $\lfloor \mu_D/2\rfloor=0$ this means a fraction $1-\alpha$ of the nodes are isolated and the remaining fraction $\alpha$ form a star). In order to compute the probability (=relative size) of an outbreak we then have to condition on if our selected index case was connected to $\lfloor \mu_D/2\rfloor$ or $\lfloor \mu_D/2\rfloor+1$ nodes. The resulting expression for the probability/size of an outbreak, also valid for the case where $\mu_D/2$ is an integer (or equivalently $\alpha=0$), is then given by:
\begin{equation}
\tau^{(epi)}_{\max}=(1-\alpha)\left(1-(1-p)^k\right) + \alpha\left(1-(1-p)^{k+1}\right),\label{tau_max_mean}
\end{equation}
where $k =\lfloor \mu_D/2\rfloor$ and $\alpha=\mu_D/2-k$.

Finally we treat the size of an epidemic outbreak in a \emph{randomly selected} network among all networks having mean degree $\mu_D$. As mentioned earlier, such a network corresponds to the Erd\H{o}s-R{\'e}nyi network and an the epidemic on the network corresponds to having the Reed-Frost epidemic model (e.g.\cite{B10,D06}) with transmission probability $p\mu_D/n$ between each pair of actors.
If an epidemic occurs on this network the final size $\tau^{(epi)}_{Rand}$ is given by the largest solution $\tau^{(epi)}_{Rand}=t$ to the equation
\begin{equation}
1-t=e^{-p\mu_Dt}.\label{tau_R-F}
\end{equation}
Note that when $p=1$ this equation coincides with Equation (\ref{tau-rand-mean}) as to be expected.
It is also known that $\tau^{(epi)}_{Rand}$ is strictly positive if and only if $p\mu_D>1$ (e.g.\cite{N03, B10}). In (\ref{tau_R-F}) it is seen that $\tau^{(epi)}_{Rand}$ only depends on the product $p\mu_D$ and not on the separate components. In Figure \ref{fig-tau-rand} we illustrate this dependence, with $\mu_D$ playing the role of $p\mu_D$.

\section{Observing egocentric data: ego only}\label{sec-deg-dist}

We now consider the case that we observe egocentric data, i.e.\ we observe the degree of a sample or all of the actors in the network. In case of a sample we neglect the uncertainty stemming from not knowing the exact degree distribution. We hence assume that we know the degree distribution $\{p_k\}$, where $p_k$ is the probability that a randomly selected actor has degree $k$.

\subsection{The giant connected component}

Just as in the previous section it is easy to construct a network consisting of small completely connected isolated units, thus achieving $\tau_{\min}=0$.
Similarly, it is possible to join all actors having degree 2 or larger into one single giant connected component by putting them in a line, actors with degree 1 can the be connected to actors having degree larger than 2 in the line. As a consequence, the size of the giant connected component is at least as large as the community fraction having degree 2 or larger, i.e.\
\begin{equation}
\tau_{\max}\ge 1-(p_0+p_1).\label{tau-max-deg}
\end{equation}
This can be made even larger by connecting the degree 1 actors to the actors which have degree larger than $2$. The mean number of actors those ``large-degree'' actors still have freedom to chose as neighbours is $\frac{\mu_D-p_1}{1-(p_0+p_1)}-2$. If this number exceeds $p_1$,  then $\tau_{\max}\ge 1-p_0$. Otherwise $\tau_{\max} = 1-(p_0+p_1) + \frac{\mu_D-p_1}{1-(p_0+p_1)}-2$.

Having solved $\tau_{\min}$ and $\tau_{\max}$ we now look at the case where we choose our network uniformly at random among all networks having degree distribution $\{p_k\}$. This is in fact exactly what is done in the configuration model (e.g.\ \cite{D06, MR, N03}) where actors are given i.i.d.\ degrees according to the degree distribution $\{p_k\}$ and edges of nodes are connected completely at random (this may of course lead to self-loops and multiple edges but it is known, \cite{D06}, that the fraction of such edges are negligible when $\mu_D<\infty$, so they may be removed without affecting the limiting degree distribution).

\begin{figure}[ht]
\begin{center}
$\begin{array}{cc}
a) \includegraphics[width=.45\textwidth]{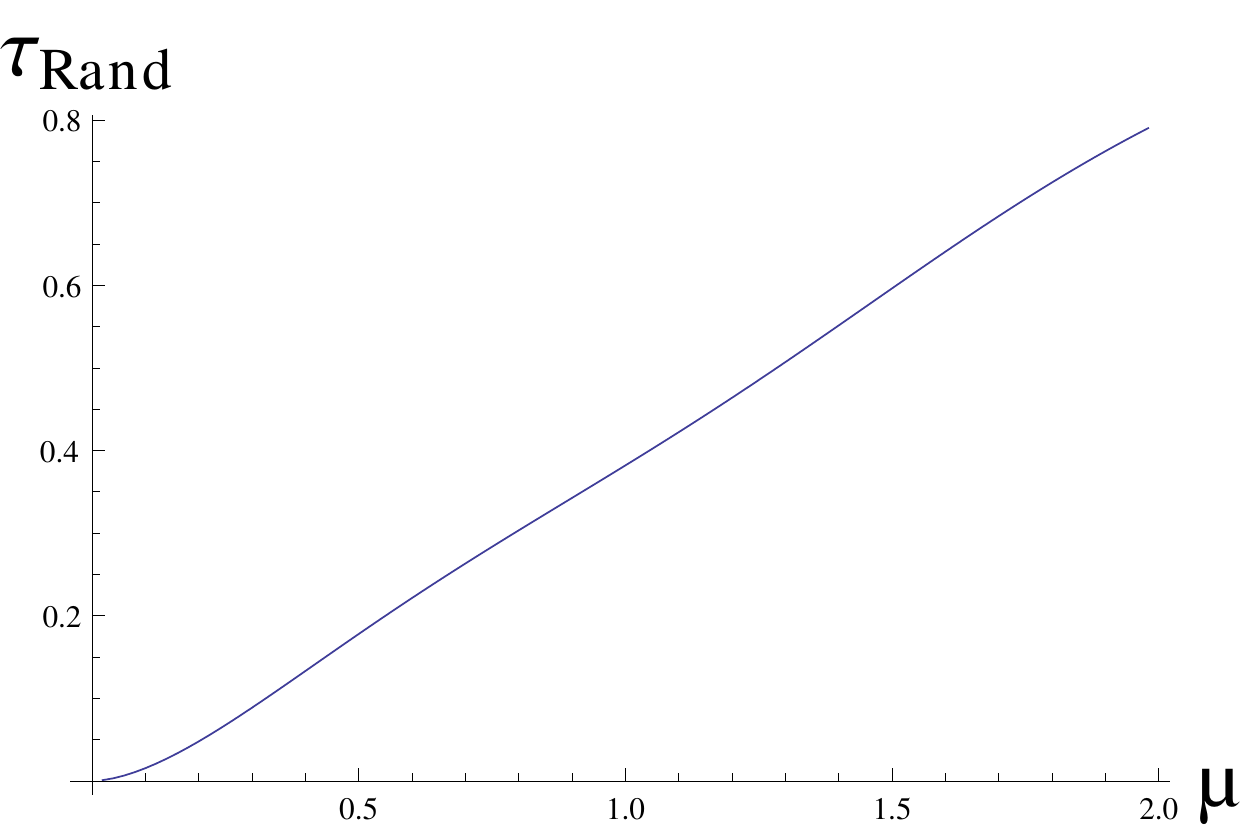}&
b) \includegraphics[width=.45\textwidth]{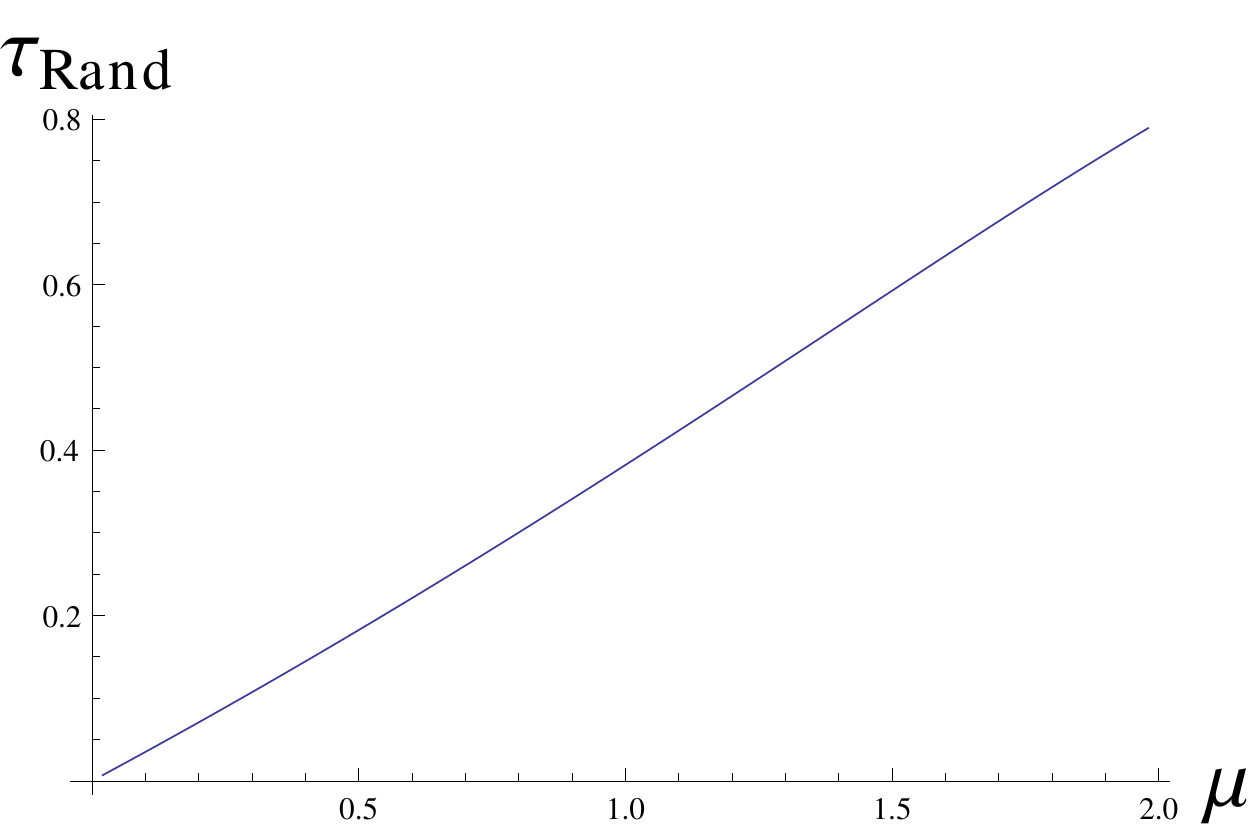}\\
c) \includegraphics[width=.45\textwidth]{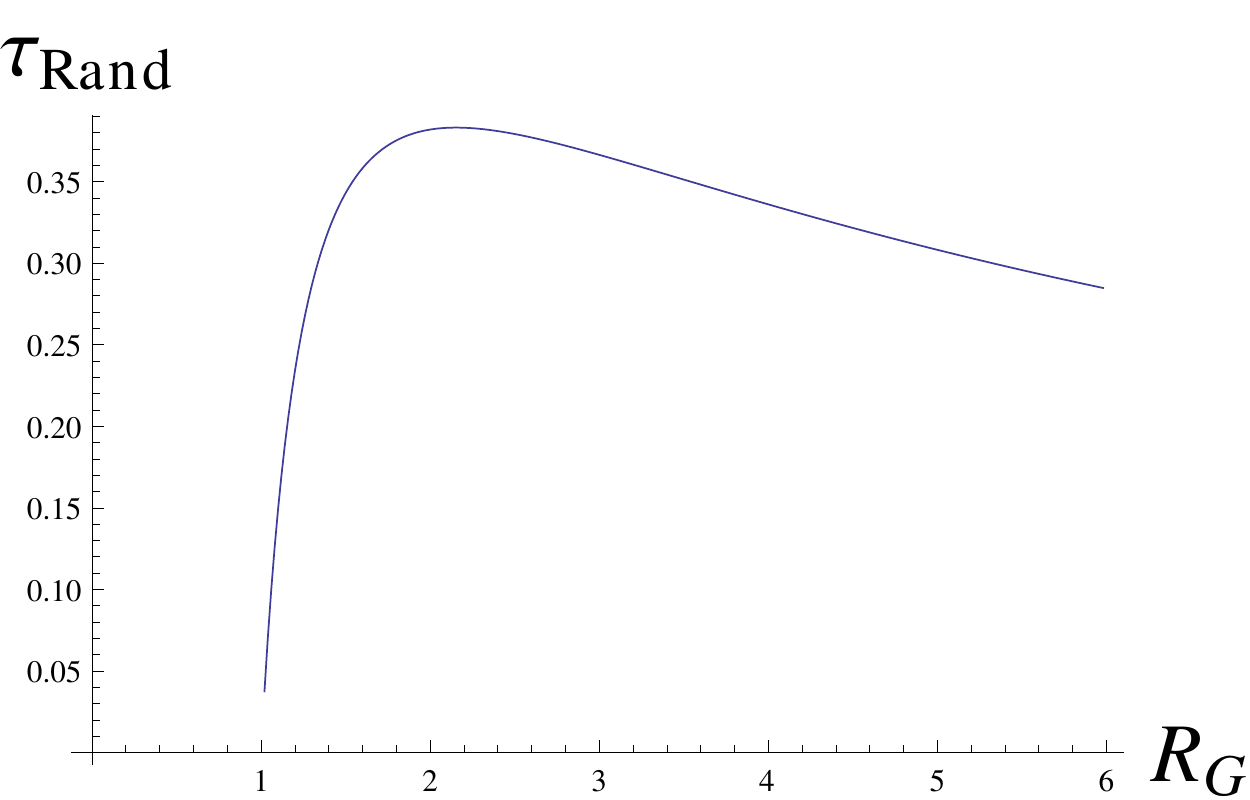} &
d) \includegraphics[width=.45\textwidth]{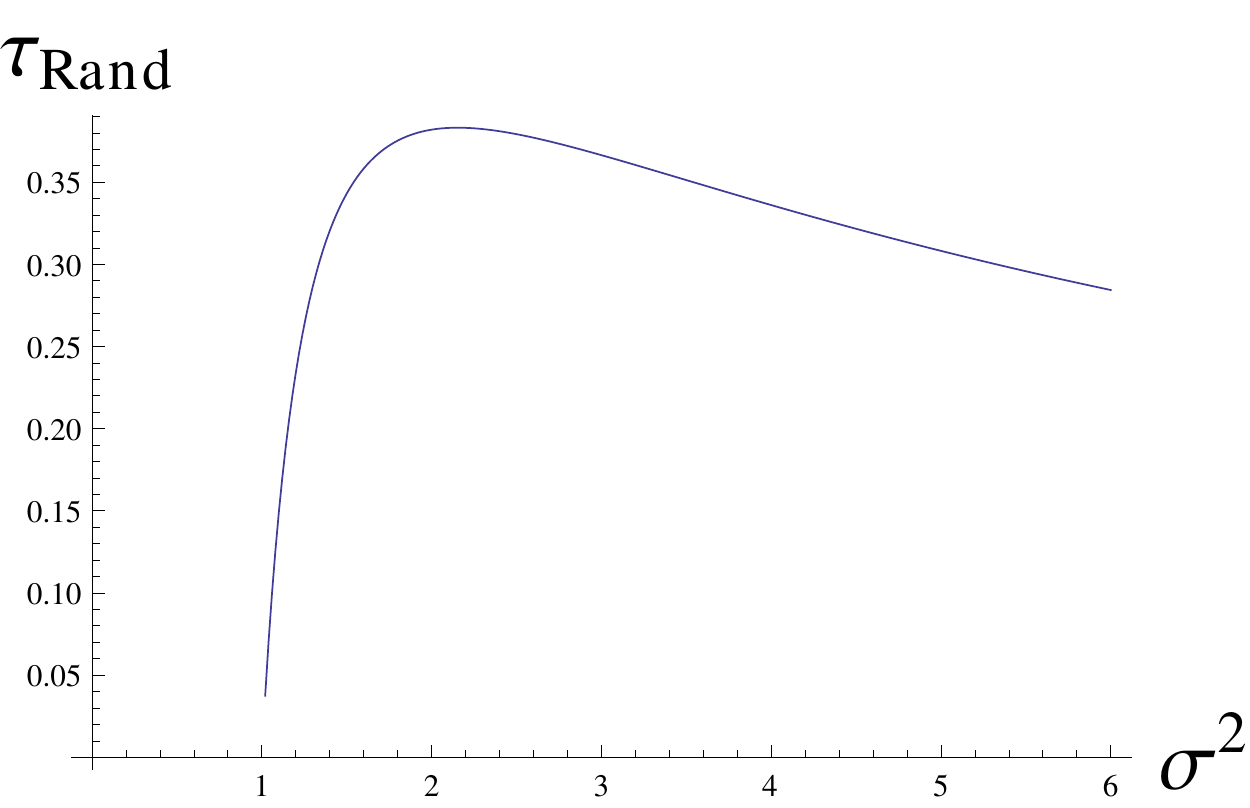} 
\end{array}$
\end{center}
\caption{Illustration of how $\tau_{Rand}$ depends on the mean $\mu_D$ of the degree distribution, with $\sigma_D^2=2$ fixed (a), with $R_G=2$ fixed (b) and how $\tau_{Rand}$ depends on $R_G$ (c) or on the variance $\sigma^2_D$ of the degree distribution (d) with $\mu_D=1$ fixed. Here $D$ has a negative binomial distribution.}\label{fig-tau-rand-deg}
\end{figure}

The relative size of the giant connected component, $\tau_{Rand}$, in a network constructed using the configuration model has already been derived (e.g. \cite{N03,D06}). Let $$\rho (s)=\sum_ks^kp_k$$ denote the probability generating function of the degree distribution and $\rho'(s)$ its derivative. Let $t=\tilde \tau$ denote the largest solution to the equation
\[
1-t=\frac{\rho'(1-t)}{\rho'(1)}.
\]
Given the solution $\tilde \tau$ (which will lie in $[0,1)$), our quantity of interest, $\tau_{Rand}$, is given by
\begin{equation}
\tau_{Rand}=1-\rho(1-\tilde \tau).\label{tau_config}
\end{equation}
It is also known that
\[\tau_{Rand}>0 \quad\text{if and only if}\quad R_G:= \mu_D+\frac{\sigma_D^2-\mu_D}{\mu_D}>1,
\]
where $\sigma_D^2$ is the variance of the degree distribution. In Figure \ref{fig-tau-rand-deg} we plot $\tau_{Rand}$ as a function of $\mu_D$ having fixed standard deviation $\sigma_D$ or $R_G$, and as a function of $\sigma_D$ or $R_G$ having fixed mean degree $\mu_D$, where $D$ has a negative binomial distribution.

The case where only the mean degree is fixed (Section \ref{sec-mean-deg}) and a randomly selected network is chosen corresponds to the case where the degree distribution is Poisson with mean $\mu_D$.

\subsection{Epidemic outbreak size}\label{sec-deg-epid}
We now look at what can happen with an epidemic (with transmission parameter $p$) occurring on a network having degree distribution $\{p_k\}$. Adding the epidemic, i.e.\ removing edges with probability $1-p$, will of course only make the size of the largest connected component smaller. So, as in the previous section, the minimal size of the largest connected component is still 0: $\tau^{(epi)}_{\min}=0$.

The corresponding maximisation problem is more involved. It is intractable to characterise how to construct a network with fixed degree distribution $\{p_k\}$ such that the epidemic outbreak size is maximal. Instead we illustrate the construction for one particular (simple) degree distribution: $p_2= 1-p_3=0.6$, i.e.\ that $60\%$ of all nodes have degree 2 and the remaining half have degree 3, implying that $\mu_D=2.4$.

The question is hence how we should connect nodes in order to maximise the size of the largest connected component in the thinned network (corresponding to the epidemic). It is obvious that we should avoid short loops because these will only reduce spreading since then some potential infectious contacts will be with already infected people. The remaining question is therefore how to connect 2-nodes (and 3-nodes respectively) to other actors. We extend the configuration model in the following way (knowing that this will result in a network without clustering). Distribute the degrees of actors as in the configuration model (i.e.\ i.i.d.\ degrees each having degree 2 with probability 0.6 and otherwise having degree 3). We now let each edge of a 2-node select an edge among the other 2-nodes with probability $r$ ($0\le r\le 1$) and with 3-nodes with the remaining probability $1-r$. In order for the total number of edges to match it follows that this implies that edges/stubs of 3-nodes should select stubs of other 3-nodes with probability $r$ as well and stubs of 2-nodes with probability $(1-r)$. The parameter $r$, which can be interpreted as the fraction of all connections to actors with the same degree, may be freely chosen in order to maximise the size of the giant. The parameter $r$  is closely related to the degree correlation: if $r$ is small we have negative degree correlation whereas if $r$ is large the degree correlation is positive.

It is straightforward to (numerically) deduce the size of an outbreak for this epidemic model. Let $\eta_2$ be the probability that an actor of degree 2, which itself is infected during the epidemic, will only generate a small number of further cases in the epidemic. Define $\eta_3$ similarly. Since an infected actor of degree 2 can infect only one other actor, which has degree 2 with probability $r$ and degree 3 with probability $1-r$, we have 
$$\eta_2 = (1-p) + pr \eta_2 + p (1-r) \eta_3.$$
Here the $1-p$ is the probability that the infected degree 2 actor will not infect other actors, while the  $pr \eta_2$ (resp.\ $p(1-r)\eta_3$) term denotes the probability that a degree 2 (resp.\ 3) actor will get infected, but does not cause many further infections.
Similarly we deduce that 
$$\eta_3 = [(1-p) + p(1-r) \eta_2 + p r \eta_3]^2.$$
From the theory on so-called branching processes \cite{J75}, we know that we need the solution for which both $\eta_2$ and $\eta_3$ are minimal.

\begin{figure}[ht]
\begin{center}
\includegraphics[width=0.6\textwidth]{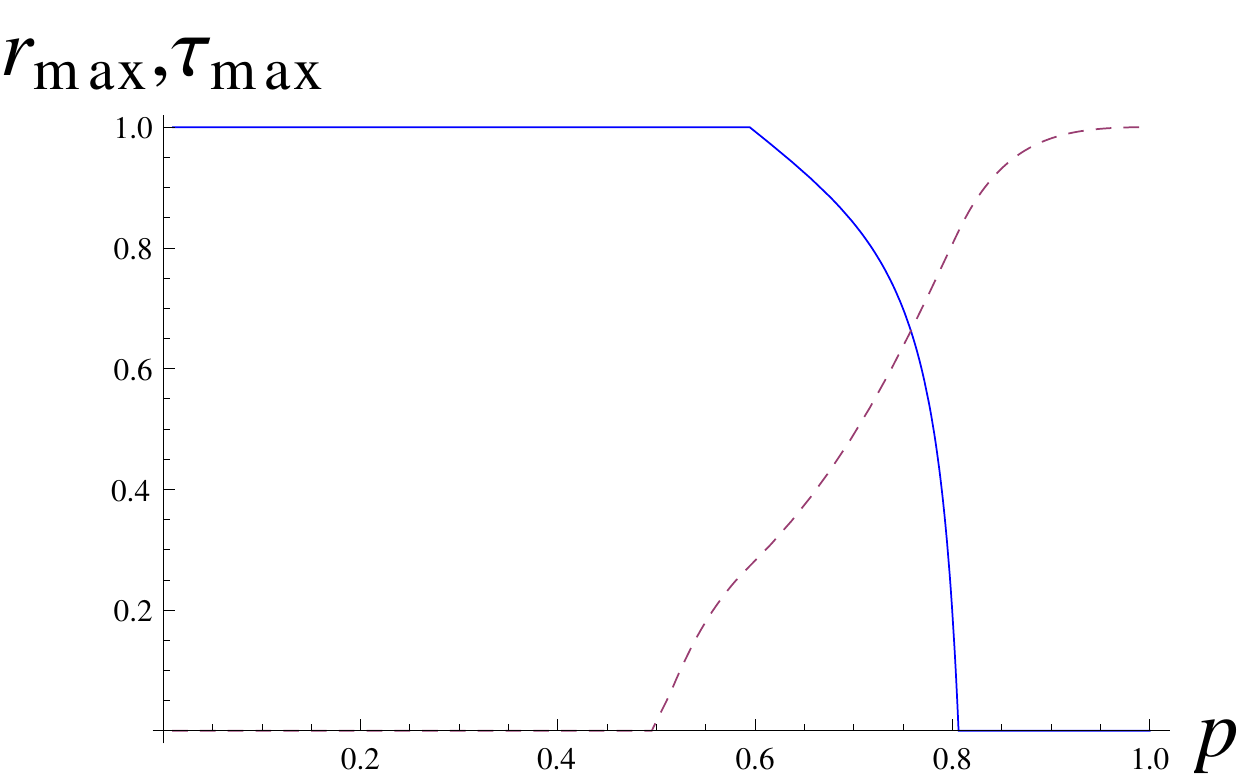}
\end{center}
\caption{The fraction $r_{max}$ of edges which connect actors of the same degree to each other, for which a large outbreak is maximised in a network in which $60\%$ of the actors has degree 2 and the other actors have degree 3, as a function of the transmission probability $p$ (solid line). The dashed line gives the corresponding relative outbreak size $\tau_{max}$ as a function of $p$.}
\label{rmax}
\end{figure}

Similar arguments give that the probability that a uniformly at random chosen actor is part of a large outbreak, if the outbreak occurs is given by
$$1- \tau^{(epi)}(\mu_D, r, p) =0.6[ (1-p) + pr \eta_2 + p (1-r) \eta_3]^2 + 0.4 [(1-p) + p(1-r) \eta_2 + p r \eta_3]^3$$
If $p\leq 1/2$, then a large outbreak has probability 0, even if all actors would have had degree 3. If on the other hand $p>1/2$, then a large outbreak is possible for some $r$. The $r$ for which the outbreak size is maximised, $r_{max}$ is given in Figure \ref{rmax}. 

The qualitative conclusion from the example, also valid for other degree distributions $\{p_k\}$, is hence that the size of the giant component $\tau^{(epi)}$ in the epidemic is maximised when nodes with high degree are connected to other nodes with high degree (and low to low) when $p$ (or more correctly $p\mu_D$) is small, and that $\tau^{(epi)}$ is maximised by the opposite construction (low to high) in the case that $p\mu_D$ is large.

Finally, the epidemic outbreak size in a randomly selected network having degree distribution $\{p_k\}$ and transmission probability $p$ is obtained exactly as for the size of the giant connected component in the randomly selected network. The only difference comes from the fact that only a binomial number of the neighbours remain connected with an actor after having thinned the network. More precisely, as has been shown in e.g.\ \cite{BJM07,D06}, $\tilde \tau$ is now the largest solution to
\[
1-t=\frac{\rho'(1-pt)}{\rho'(1)}.
\]
And, given the solution $\tilde \tau$, our quantity of interest, $\tau^{(epi)}_{Rand}$, is then given by
\begin{equation}
\tau^{(epi)}_{Rand}=1-\rho(1-p\tilde \tau).\label{tau_config_epid}
\end{equation}
Similar to before, it also holds that
\[
\tau^{(epi)}_{Rand}>0 \quad\text{if and only if}\quad R_0=p\left(\mu_D+\frac{\sigma_D^2-\mu_D}{\mu_D}\right) >1,
\]
where $R_0$ denotes the basic reproduction number.

If we know that a network is well described by a configuration model and we know the mean degree of the actors, $\mu_D$, One further question to answer is: for which distribution $\{p_k\}$ is the size of a large outbreak maximised when the transmission probability equals $p$? In \cite{BT} it is shown that the answer to this question depends on $p$ and $\mu_D$, but in all cases  the degree distribution should be non-zero at at most 2 consecutive positive integer numbers and possible at degree $0$. 

\section{Observing egocentric data with alter connections}
We end our analysis with the situation where the number of connections of all (or a sample of) actors are observed, and where it is also observed which of the connections of an actor are themselves connected. Such data, referred to as egocentric data with alter connections (e.g.\ \cite{HR}), may often be collected in egocentric network surveys since egos are usually aware of this information.

Such data gives the degree distribution in the community, but also, for each degree, the community frequency of having any given set of fully connected components of various sizes. As an example, one would know what fraction of the community that have degree 6 where two connections are not connected to any other connection, and the remaining four are connected pairwise (forming 2 triangles with ego). As mentioned previously we simplify this type of data to knowing only the degree distribution and how many of the connections are not connected with others and how many are connected pairwise, with the implicit assumption that having larger fully connected components than triangles is unlikely (cf.\ \cite{M09}). The distribution is hence specified by $\{p(k_1,k_{\Delta})\}$, where $p(k_1,k_{\Delta})$ is the probability that a randomly selected ego has $k_1$ connections that are not connected to other connections of ego, and $k_{\Delta}$ \emph{pairs} of connections that are also connected themselves pairwise. The total degree of such an actor is hence $k_1+2k_{\Delta}$.

\subsection{The giant connected component}
\begin{figure}[ht]
\begin{center}
\includegraphics[width=0.4\textwidth]{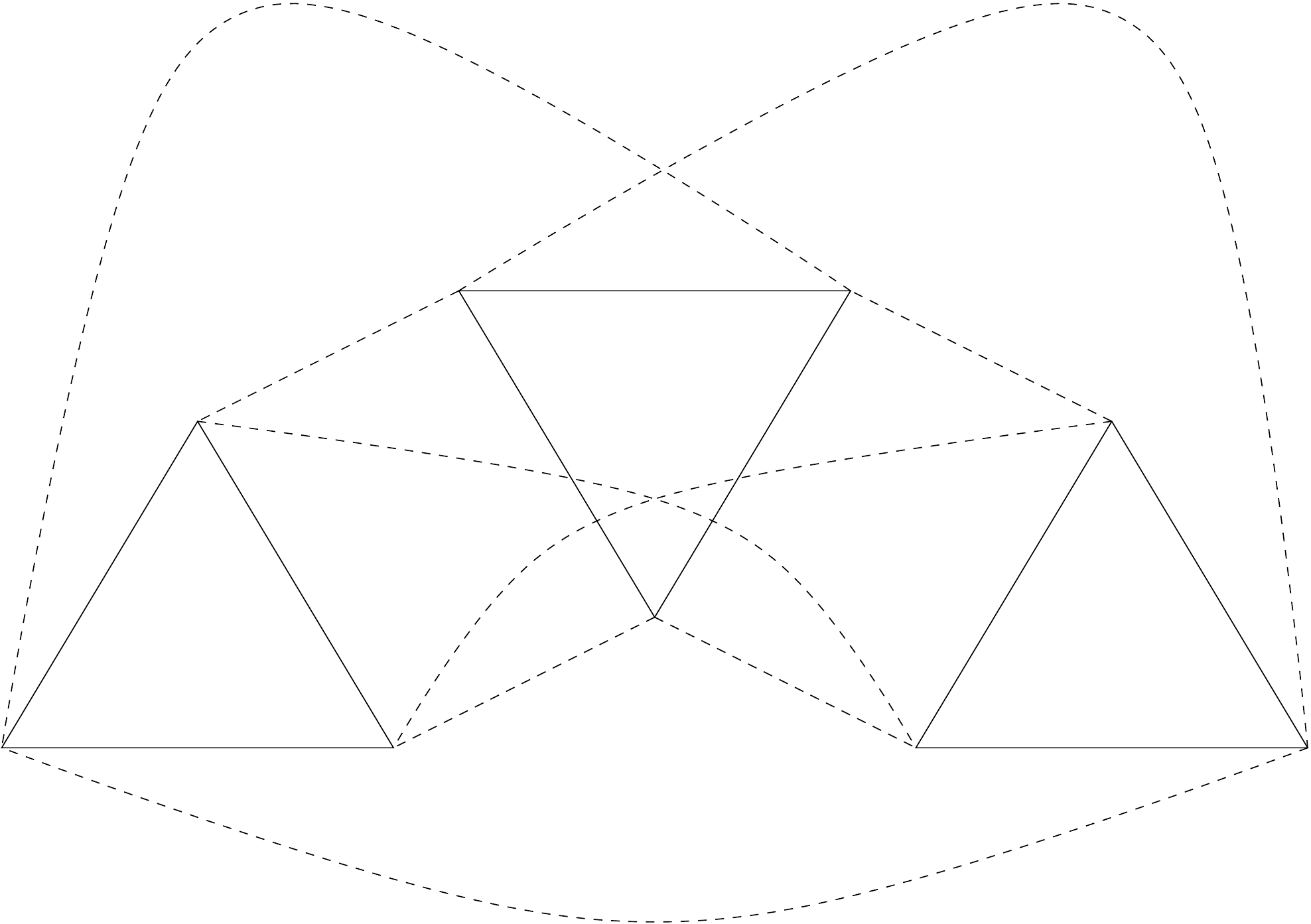}
\end{center}
\caption{A component of a graph where all elements have degree pair $(k_1=2,k_{\Delta}=1)$}
\label{messygraph}
\end{figure}

As in the previous situations it is possible to construct a network consisting of only small connected components. Now that the number of triangles each actor belongs to is pre-specified, this is a bit more involved as it is no longer possible to join egos into fully connected components. However, it is possible to pick suitably many egos of a given degree pair $(k_1,k_{\Delta})$ such that they can form an isolated component; for $(k_1=2,k_{\Delta}=1)$ it suffices with 9 egos to form an isolated component, (see Figure \ref{messygraph}). As a consequence, it is possible to construct a network without a giant component, so $\tau_{\min}=0$.

Similarly, it is in most situations, possible to construct a network in which all egos are connected (i.e.\ $\tau_{\max}=1$). This may not be the case if the degrees (of both sorts) are too small; then some egos have to be ''sacrificed'' just like before. We will not characterise which degree distributions that allow for all egos being connected (i.e.\ $\tau_{\max}=1$) and how large the giant may be if this is not the case.

Now to the relative size of the largest connected component of a \emph{random} network having the specified distribution $\{ p(k_1,k_{\Delta})\}$ of singleton neighbours and pairs of interconnected neighbours. For this we use results by Miller \cite{M09} who derives $\tau_{Rand}$ for such a random network. The recipe is given in the next subsection for the special case where the transmission probability $p$ equals 1.

\subsection{Epidemic outbreak size}

Removing edges due to no transmission will never increase the size of the giant component, so we still have $\tau^{(epi)}_{\min}=0$ as for the case without the epidemic.

Next we present how to derive the relative size of the giant of a \emph{randomly selected} network having the prescribed degree and triangle distribution $\{ p(k_1,k_{\Delta})\}$ using methods from \cite{M09}. This is done by first solving 4 unknowns $g_1,\ g_{\Delta},\ h_1,\ h_{\Delta}$ from 4 equations. The interpretation of $g_1$ and $\ g_{\Delta},$ are as the probability that a singleton edge, or triangle respectively, of a randomly selected node does \emph{not} connect to the giant component, and $h_1$ and $h_{\Delta}$ are the probabilities that a node reached by a randomly selected singleton edge, or triangle respectively, does not connected to the giant from this edge/triangle. The four equations are:
\begin{align*}
g_1&=1-p+ph_1,\\
h_1&= \frac{1}{E(D_1)}\sum_{k_1,k_{\Delta}}k_1 p(k_1,k_{\Delta})g_1^{k_1-1} g_{\Delta}^{k_{\Delta}},\\
g_{\Delta}&= (1-p+ph_{\Delta})^2-2p^2(1-p)h_{\Delta}(1-h_{\Delta}),\\
h_{\Delta}&= \frac{1}{E(D_{\Delta})} \sum_{k_1,k_{\Delta} }k_{\Delta} p(k_1,k_{\Delta})g_1^{k_1} g_{\Delta}^{k_{\Delta}-1}.
\end{align*}
These equations can be solved iteratively beginning with e.g.\ $h_1=h_{\Delta}=0$ thus giving the numerical solutions $g_1,\ h_1,\ g_{\Delta},\ h_{\Delta}$. Given these solutions we have that the relative final size of a major epidemic outbreak in a random network with single- and triangle-degree distribution $\{ p(k_1,k_{\Delta})\}$ and with transmission probability $p$ is given by
\begin{equation}
\tau^{(epi)}_{Rand}=1- \sum_{k_1,k_{\Delta}} p(k_1,k_{\Delta})g_1^{k_1} g_{\Delta}^{k_{\Delta}}.\label{tau_rand_tri}
\end{equation}
Further, $\tau^{(epi)}_{Rand}$ is  strictly positive if and only if the basic reproduction number $R_0$ exceeds the value of 1, and it is shown in \cite{M09} that $R_0$ is the dominant eigenvalue of the $2\times 2$-matrix  $M$ defined by
\[M =\left(
\begin{array}{cc}
\frac{pE(D_1^2-D_1)}{E(D_1)} & \frac{pE(D_1D_{\Delta})}{E(D_{\Delta})}\\
\frac{2p(1+p-p^2)E(D_1D_{\Delta})}{E(D_1)} & \frac{2p(1+p-p^2)E(D_{\Delta}^2-D_{\Delta})}{E(D_{\Delta})}
\end{array}
\right).\]

The corresponding result for the giant component of the original network is obtained by setting $p=1$ in the equations above which reduces the number of equations to be solved iteratively from 4 down to 2.

Having derived $\tau^{(epi)}_{Rand}$ defined in (\ref{tau_rand_tri}) we can as before ask: is it possible to have a bigger epidemic outbreak than $\tau_{Rand}$ for fixed degree distribution $\{ p(k_1,k_{\Delta})\}$ and transmission probability $p$. The answer is ''yes'', as it was for the case with given singleton degree distribution and no triangles (cf.\ Section \ref{sec-deg-epid}). In fact, a larger outbreak is possible to obtain if large-degree egos are connected to other large degree ego in the case that the transmission probability and mean degrees are small, and by connecting large-degree egos to small-degree egos when these quantities are large. To try to characterise exactly how this should be done for a given degree distribution $\{ p(k_1,k_{\Delta})\}$ is however not very instructive and is hence omitted.

\begin{figure}[ht]
\begin{center}
\includegraphics[width=.5\textwidth]{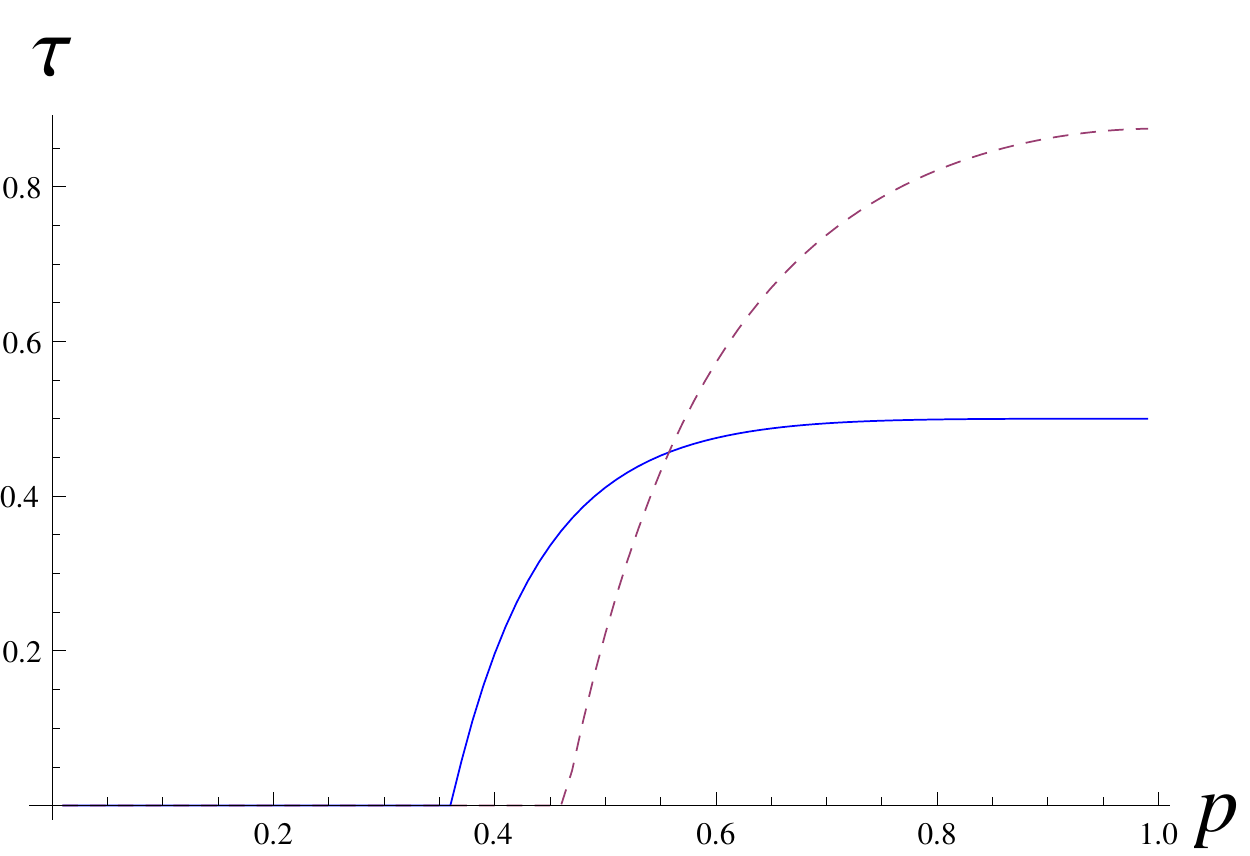}
\end{center}
\caption{Illustration of how $\tau=\tau^{(epi)}$ varies with $p$ for a given independent and triangle distribution, both the random case (dashed line) and the case where nodes of similar degree tend to be connected (solid line).}\label{fig-tau-rand-deg-clust}
\end{figure}

In Figure \ref{fig-tau-rand-deg-clust} we illustrate how the relative size of the giant component varies with the transmission probability $p$ for the case where $p(k_1=0,k_{\Delta}=1)=p(k_1=2,k_{\Delta}=1)=0.5$, i.e.\ where all actors belong to one triangle and half of the actors have no other connections and the other half have two independent singleton edges on top of this. We plot both the case where the triangles are connected completely at random (zero degree correlation) and the extreme where triangles are always formed by connecting actors having the same degree. It is seen that positive degree correlation gives the largest outbreak size when $p$ is small whereas zero degree correlation gives larger outbreak size when $p$ is close to 1.

\section{Discussion}

In the paper it was described how large/small the giant connected component, as well as the size of an epidemic outbreak occurring on the social network, might be for some given information about the egocentric network. For all types of egocentric data it is possible not to have a giant (or epidemic outbreak) of the same order as the network. However, the upper bound on the size of the giant/outbreak decreases the more detailed egocentric data is available. For the epidemic case, a larger outbreak than that of a randomly selected network among those consistent with the egocentric data, is obtained by connecting actors with high degree to low degree actors if the transmission probability $p$ and the degrees are large, and to connected actors with high degree to other actors if these numbers instead are small. That is, if the degrees and transmission probability are large, then we get a larger outbreak if the degree correlation is negative, and if the degree and transmission probability are small we get a larger outbreak if the degree correlation is positive.

In the data form denoted egocentric with alter connection it was assumed that actors only had neighbours that were not connected to any other neighbour of ego, or else that were connected to exactly one other neighbour of ego. This is of course a simplification of real world networks (for example household are usually treated as a fully connected group of actors). It is an open question to see what effect such a deviation from the model assumption has on the network properties.

In the paper we studied three different levels of detailed egocentric data: mean degree, degree distribution, and degree distribution including singleton and triangle degree. The only global properties treated were the size of the giant and of a possible epidemic outbreak in the community. There are many other global properties worthy of analysis under the same scenario, for example the diameter and betweeness of the network.

\section*{Acknowledgements}
T.B. is grateful to
Riksbankens Jubileumsfond for financial support.
P.T.\ is supported through Vetenskapsr{\aa}det (Swedish research counsel) projectnr.\ 2010--5873.

\end{document}